\documentclass[aps,prd,twocolumn,superscriptaddress,nofootinbib,floatfix,noshowpacs]{revtex4-1}
\pdfoutput=1
\newif\ifcomment
\usepackage[T1]{fontenc}
\usepackage{amsmath,amssymb}
\usepackage{color,hyperref,url}
\usepackage{listings}
\usepackage{slashed}
\usepackage[pdftex]{graphicx}
\usepackage{epstopdf}
\usepackage{epsfig}
\usepackage{grffile}
\usepackage{relsize}
\graphicspath{{./img/}}
\usepackage{soul}
\usepackage{multirow}
\usepackage{adjustbox}

\newcommand{\al}{\alpha}
\newcommand{\be}{\beta}

\newcommand{\beq}{\begin{equation}}
\newcommand{\eeq}{\end{equation}}
\newcommand{\ba}{\begin{array}}
\newcommand{\ea}{\end{array}}
\newcommand{\bea}{\begin{align}}
\newcommand{\eea}{\end{align}}
\newcommand{\bi}{\begin{itemize}}
\newcommand{\ei}{\end{itemize}}
\newcommand{\ben}{\begin{enumerate}}
\newcommand{\een}{\end{enumerate}}
\newcommand{\bc}{\begin{center}}
\newcommand{\ec}{\end{center}}
\newcommand{\bl}{\begin{flushleft}}
\newcommand{\el}{\end{flushleft}}
\newcommand{\br}{\begin{flushright}}
\newcommand{\er}{\end{flushright}}

\newcommand{\nn}{\nonumber \\}
\newcommand\Eqn[1]{Eq.~(\ref{#1})}  

\newcommand{\ie}{{i.e.}}

\begin{document}
\title{Electromagnetic structure of axial-vector mesons and implications for the muon $g-2$}

\author{Zanbin Xing}
\email{xingzb@mail.nankai.edu.cn}
\affiliation{School of Physics, Nankai University, Tianjin 300071, China}

\author{Lei Chang}
\email{leichang@nankai.edu.cn }
\affiliation{School of Physics, Nankai University, Tianjin 300071, China}

\author{Kh\'epani Raya}
\email{khepani.raya@dci.uhu.es}
\affiliation{Department of Integrated Sciences and Center for Advanced Studies in Physics, Mathematics and Computation, University of Huelva, E-21071 Huelva, Spain.}

\date{\today}
\begin{abstract}
The electromagnetic structure of axial-vector mesons is investigated via elastic and two-photon transition form factors (TFFs). To this end, we employ a framework based on the Dyson-Schwinger and Bethe-Salpeter equations  within a contact interaction model. This largely algebraic approach transparently exposes the role of symmetries and their breaking, and has proven successful in describing anomaly-sensitive processes, including pseudoscalar to two-photon TFFs, $\gamma \to 3 \pi$, and vector-to-pseudoscalar radiative decays.  Restricting our analysis to the lowest-lying states, $\text{A}=\{ a_1, f_1,f_1'\}$, we also evaluate the corresponding light-by-light contribution to the muon anomalous magnetic moment, and obtain $a_\mu^{\text{A}}=11.30(4.71)\times10^{-11}$, consistent with contemporary estimates.
\end{abstract}

\maketitle
\section{Introduction}

The anomalous magnetic moment of the muon, commonly dubbed muon $g-2$ or simply $a_\mu$, remains one of the most compelling observables in modern physics. Following the most recent analysis of the Muon g‑2 experiment at Fermilab, the empirical value of this quantity has been determined as 
$a_\mu^{\text{Exp}}=116592071.5(14.5)\times10^{-11}$\,\cite{Muong-2:2025xyk}. Meanwhile, the Muon g‑2 Theory Initiative has reached a new consensus, establishing $a_\mu^{\text{SM}}=116592033(62)\times10^{-11}$ as the Standard Model (SM) prediction\,\cite{Aliberti:2025beg}. Contrary to the situation that persisted for decades, these latest values suggest that experiment and SM are not in conflict. However, neither for the new physics enthusiasts nor for the most conservative, this is not the end of the story. The intricate nature of the nuclear strong interactions, described within the SM by quantum chromodynamics (QCD), poses a huge challenge in determining the hadronic contributions to $a_\mu$\,\cite{Aliberti:2025beg,Aoyama:2020ynm,Colangelo:2022jxc}. These components, divided into hadronic vacuum polarization (HVP) and hadronic light-by-light (HLbL), dominate the uncertainty in $a_\mu^{\text{SM}}$, preventing its theoretical prediction from reaching the same level of precision as its experimental counterpart. Axial-vector (AV) mesons, while subdominant compared to pseudoscalars, remain a key source of uncertainty in the HLbL sector, to a great extent because of the limited experimental knowledge of their relevant transition form factors (TFFs)\,\cite{Szczurek:2020hpc,Hoferichter:2023tgp,Zanke:2021wiq,Roig:2019reh}. Axial-vector mesons are also essential for enforcing short-distance QCD constraints in HLbL scattering\,\cite{Melnikov:2003xd,Masjuan:2020jsf}. This work is therefore devoted to the study of the two-photon to AV meson TFFs and their contribution to $a_\mu$.

It is well-kown that the construction of the AV to two-photon transition amplitude presents a number of challenges\,\cite{Kuhn:1979bb,Rudenko:2017bel,Roig:2019reh,Masjuan:2020jsf,Hoferichter:2020lap,Eichmann:2024glq}. First, as a spin-1 hadron, its Lorentz structure is naturally more intricate than in the pseudoscalar case. Symmetry requirements and gauge invariance, among others, add to this complexity. Moreover, the Landau–Yang theorem forbids the decay of a spin-1 particle into two on-shell photons,\,\cite{Landau:1948kw,Yang:1950rg}. Consequently, while the on-shell amplitude vanishes, AV meson TFFs are probed and defined for configurations with at least one off-shell photon, where the choice of tensor basis is no longer irrelevant\,\cite{Roig:2019reh}. Because of this limitation, experimental extractions of these form factors remain challenging, and robust theoretical descriptions become indispensable. Furthermore, while vector mesons have been extensively investigated (e.g. Refs.\,\cite{Dudek:2006ej,Owen:2015gva,Xu:2019ilh,Xing:2021dwe,Almeida-Zamora:2023rwg,Rojas:2024tmn,Serna:2020txe,Hernandez-Pinto:2024kwg,Xu:2025sxw}), axial-vector mesons remain much less explored. The analysis of the structural similarities and differences between these correlated systems, i.e. parity partners, is particularly valuable for probing dynamical chiral symmetry breaking (DCSB). This mechanism strongly shapes the internal structure of hadrons\,\cite{Raya:2024ejx,Roberts:2021nhw}, making its proper understanding essential for accurate Standard Model predictions. In that regard, the exploration of the electromagnetic structure of the AV mesons, as revealed by the corresponding elastic electromagnetic form factors (EFFs), also deserves consideration.

Our approach to QCD is based upon a combination of the Dyson–Schwinger (DS) and Bethe–Salpeter (BS) equations\,\cite{Roberts:1994dr,Eichmann:2016yit,Qin:2020rad}. This formalism captures essential characteristics of QCD, such as confinement and DCSB, and has be applied to a wide range of quantities characterizing the hadronic structure, including distribution functions and form factors (see e.g.\,\cite{Raya:2024ejx,Roberts:2021nhw}). Concerning $a_\mu$, both pseudoscalar pole and box contributions have been investigated\,\cite{Raya:2019dnh,Miramontes:2021exi,Miramontes:2024fgo,Eichmann:2019tjk,Eichmann:2019bqf}. More recently, the axial-vector and scalar mesons contributions were discussed in Ref.\,\cite{Eichmann:2024glq}. Rather than relying on a numerically intensive computation, we employ a primarily algebraic framework that makes the origin of symmetries and anomalies transparent. Specifically, we work within the contact interaction (CI) model presented in Ref.\,\cite{Xing:2021dwe}, which has proven successful in describing anomaly-sensitive processes such as $\{\pi, \eta, \eta'\} \to \gamma\gamma$, $\gamma \to 3\pi$, and vector-to-pseudoscalar radiative decays\,\cite{Dang:2023ysl,Xing:2024bpj,Xu:2024frc}.  Note from a DS-BS equations perspective, the description of AV mesons is rather challenging. Among other issues, typical truncations such as the well-known rainbow–ladder (RL) approximation underestimate the mass splittings between parity partners, with the $m_{a_1}-m_\rho$ splitting being a prime example\,\cite{Lu:2017cln,Yin:2021uom,Gutierrez-Guerrero:2021rsx,Qin:2020jig,Chang:2009zb}. This shortcoming largely arises from the absence of certain effects in the interaction kernels, such as spin–orbit repulsion. In addition to systematic rescalings of the RL results,\,\cite{Eichmann:2024glq}, improved kernels have been proposed\,\cite{Qin:2020jig,Chang:2009zb,Chang:2021vvx}. Nonetheless, as anticipated, here we adopt a simpler and more transparent approach to clarify the role of the AV meson mass in the elastic and transition form factors, and its implications for $a_\mu$.

The manuscript is organized as follows. We first present the employed formalism, including the definitions of the matrix elements, their associated form factors, and the essentials of their computation within the CI model. Next, we provide numerical results for the produced mass spectrum, elastic and transition form factors, and the corresponding contribution $a_\mu$. We capitalize on the role of the AV meson mass on determining these quantities. Finally, we summarize our findings and outline directions for future related works.

\section{Formalism}
\label{sec:Formalism}

\subsection{Matrix elements and form factors}
\label{sec:FormalismTFFs}

Let us first recall the matrix element that characterizes the coupling of a photon to an axial-vector meson\,\cite{Raya:2021pyr}:
\begin{eqnarray}
\label{eq:lambdavertex1}
\Lambda_{\mu\nu\rho}(K,Q)&=&\sum_{j=1}^3 T^{j}_{\mu\nu\rho}(K,Q) G_j(Q^2) \;\\\label{eq:lambdavertex2}
T^{1}_{\mu\nu\rho}(K,Q)&=&2K_\mu P_{\nu\alpha}^T(p_i)P_{\alpha\rho}^T(p_f)\;,\\ \nonumber
T^{2}_{\mu\nu\rho}(K,Q)&=&\left[Q_\nu-p_{\nu}^i\frac{Q^2}{2m_{\text{A}}^2}\right]P_{\mu\rho}^T(p_f) \nonumber \\
&-&\left[Q_\rho+p_{\rho}^f\frac{Q^2}{2m_{\text{A}}^2}\right]P_{\mu\nu}^T(p_i)\;,\\ \nonumber
T^{3}_{\mu\nu\rho}(K,Q)&=&\frac{K_\mu}{m_{\text{A}}^2}\left[Q_\nu-p_{\nu}^i\frac{Q^2}{2m_{\text{A}}^2}\right]\left[Q_\rho+p_{\rho}^f\frac{Q^2}{2m_{\text{A}}^2}\right]\;.
\end{eqnarray}
Here $P_{\mu\nu}^T(p)=\delta_{\mu\nu}-p_\mu p_\nu/p^2$ and the kinematics is defined as follows: $p_i = K-Q/2$  and $p_f = K+Q/2$ denote the incoming and outgoing meson momenta, respectively, and $Q=p_f-p_i$ is the photon momentum. The on-shell conditions, $p_i^2=p_f^2=-m_{\text{A}}^2$ ($m_{\text{A}}$ the mass of the AV meson), imply $K\cdot Q=0$ and $K^2=-m_{\text{A}}^2-Q^2/4$. In addition, the preservation of Ward-Green-Takahashi identities impose\,\cite{Xing:2021dwe}:
\begin{eqnarray}
p_\nu^i \Lambda_{\mu\nu\rho} = p_\rho^f\Lambda_{\mu\nu\rho} = Q_\mu \Lambda_{\mu\nu\rho}=0\;.
\end{eqnarray}
The electric ($G_E$), magnetic ($G_M$) and quadrupole ($G_Q$) form factors can be constructed in terms of $G_{1,2,3}(Q^2)$ as follows ($\eta=Q^2/(4m_\text{A}^2)$):
\begin{subequations} \label{eq:GEMQ}
\begin{align}
G_E(Q^2)=&\,G_1(Q^2)+\frac{2}{3}\eta \,G_Q(Q^2)\,, \label{eq:GEMQ1} \\
G_M(Q^2)=&\,-G_2(Q^2)\,,\label{eq:GEMQ2} \\
G_Q(Q^2)=&\,G_1(Q^2)+G_2(Q^2)+G_3(Q^2)\left[1+\eta\right]\,.\label{eq:GEMQ3}
\end{align}
\end{subequations}

On its part, the AV meson to two-photons transition amplitude, $\mathcal{A}_{\mu\nu\rho}$, can be parameterized as:
\begin{equation}
\label{eq:defTA}
    \mathcal{A}_{\mu\nu\rho}(Q,Q')=\sum_{i}T^i_{\mu\nu\rho}(Q,Q') f_i(Q^2,Q'^2)\,,
\end{equation}
where $Q,Q'$ are the photon momenta, and $f_i$ are Lorentz scalar functions. The tensors $T^i$ might be expressed rather generally as follows\,\cite{Eichmann:2024glq}:
\begin{eqnarray}
T_{\mu\nu\rho}^1=\epsilon_{\mu  \nu  \rho  Q}&,&T_{\mu\nu\rho}^2=\epsilon_{\mu  \nu  \rho  Q'}\,,\nn
T_{\mu\nu\rho}^3=Q_{\mu }
\epsilon_{\nu  \rho  Q
   Q'}&,&
T_{\mu\nu\rho}^4=Q'_{\mu } \epsilon_{\nu  \rho  Q Q'}\,,\nn
T_{\mu\nu\rho}^5=Q_{\nu }\epsilon_{\mu  \rho  Q
   Q'}&,&
T_{\mu\nu\rho}^6=Q'_{\nu }\epsilon_{\mu  \rho  Q Q'}\,,\nn
T_{\mu\nu\rho}^7=Q_{\rho }\epsilon_{\mu  \nu  Q
   Q'}&,&
T_{\mu\nu\rho}^8=Q'_{\rho }\epsilon_{\mu  \nu  Q Q'}\,.
\label{eq:OurTis}
\end{eqnarray}
The momentum appearing in the index of Levi-Civita symbol is understood as a contraction. Note that the last two structures can be written in terms of the rest according to the Schouten identities ($X=\{Q,\,Q'\}$):
\begin{align}
    -X_\rho \epsilon_{QQ'\mu\nu}&=X_\mu \epsilon_{\nu\rho QQ'}+X_\nu \epsilon_{\rho QQ' \mu}\nn&+X\cdot Q \epsilon_{Q' \mu\nu\rho}+X\cdot Q' \epsilon_{\mu\nu\rho Q}\,.
\end{align}
Bose-symmetry implies $\mathcal{A}_{\mu\nu\rho}(Q,Q')=\mathcal{A}_{\nu\mu\rho}(-Q',-Q)$ and, according to the gauge invariance and transversality of the AV meson,
\begin{equation}
\label{eq:WGTI1}
    Q_\mu \mathcal{A}_{\mu\nu\rho}=0\,,\,
    Q'_\nu \mathcal{A}_{\mu\nu\rho}=0\,,\,
    P_\rho \mathcal{A}_{\mu\nu\rho}=0\,.
\end{equation}
Here $P=Q-Q'$, such that the on-shell condition $P^2=-m_{\text{A}}^2$ entails $2Q\cdot Q'=Q^2+Q'^2+m_{\text{A}}^2$. The constraints in Eq.\,\eqref{eq:WGTI1} relate the scalar functions $f_i$ as follows:
\begin{subequations} \label{eq:wtisf}
\begin{align}
f_2+Q^2 f_3+Q\cdot Q' f_4=0&\,, \label{eq:wtis1} \\
f_1+Q\cdot Q' f_5+Q'^2 f_6=0&\,,\label{eq:wtis1e} \\
f_1+f_2+Q\cdot P f_7+Q'\cdot P f_8=0&\,.\label{eq:wtis2}
\end{align}
\end{subequations}

A consequence of these identities is that, at the end, there are only three independent Lorentz structures and corresponding scalar functions for the amplitude $\mathcal{A}_{\mu\nu\rho}$. Accordingly, we adopt the form factors $F_{1,2,3}$ defined in Ref.\,\cite{Eichmann:2024glq}. The associated tensor basis is detailed therein and provided here in Appendix\,\ref{ap:altbasis}. Its connection to scalar functions $f_i$ is as follows:
\begin{subequations} \label{eq:Fis}
\begin{align}
F_1(Q^2,Q'^2)=&(f_3+f_4+f_5+f_6)\,m_{\text{A}}^2\,, \label{eq:Fis1} \\
F_2(Q^2,Q'^2)=&\frac{2m_{\text{A}}^4}{Q'^2-Q^2}(f_3-f_4+f_5-f_6)\,,\label{eq:Fis2} \\
 F_3(Q^2,Q'^2)=&\frac{m_{\text{A}}^4}{Q'^2-Q^2}(f_4-f_5-f_7-f_8)\,.\label{eq:Fis3}
\end{align}
\end{subequations}

Having established the structure of $\Lambda_{\mu\nu\rho}$ and $\mathcal{A}_{\mu\nu\rho}$, we now discuss the evaluation of the corresponding form factors within our symmetry-preserving scheme.

\subsection{Impulse approximation}
We shall consider the so called impulse approximation to evaluate the $\Lambda_{\mu\nu\rho}$, in whose case:
\begin{eqnarray}
\label{eq:impulse}
\Lambda_{\mu\nu\rho}(Q^2)&=&2 N_c\ \text{tr}_D\int_l \Gamma_\rho^{\text{A}}(-p_f)S(l+p_f)\nonumber\\ 
&\times& i\Gamma_\mu^\gamma(Q)S(l+p_i)\Gamma_\nu^{\text{A}}(p_i)S(l)\;.
\end{eqnarray} 
An analogous expression holds for the transition amplitude $\mathcal{A}_{\mu\nu\rho}$:
\begin{eqnarray}
\label{eq:IA}
\mathcal{A}_{\mu\nu\rho}&=&2N_{cf}\text{tr}_D\int_l S\left(l+\frac{P}{2}\right)\Gamma^{\text{A}}_\rho(P)S\left(l-\frac{P}{2}\right) \nn
&\times& \Gamma^\gamma_\mu(-Q) S\left(l+\frac{Q+Q'}{2}\right) \Gamma^\gamma_\nu(Q')\,.
\end{eqnarray}
Here $\int_l:=\int d^4l/(2\pi)^4$ is a Poincaré invariant four-momentum integral. The symbol $\text{tr}_D$ indicates a trace over Dirac space, $N_c=3$ is the number of colors, and $N_{cf}$ arises from the color and flavor traces, such that:
\begin{equation}
\label{eq:flavorNorm}
    N_{cf}^{a_1}=\frac{1}{\sqrt{2}}\,,\,
    N_{cf}^{f_1}=\frac{5}{3\sqrt{2}}\,,\,N_{cf}^{f_1'}=\frac{1}{3}\,.
\end{equation}
The rest of the components in Eqs.\,\eqref{eq:impulse} and \eqref{eq:IA} are defined as usual: $S(p)$ is the fully-dressed quark propagator, while $\Gamma_\rho^{\text{A}}$  and $\Gamma_{\mu,\nu}^\gamma$ denote the AV meson's Bethe-Salpeter amplitude (BSA) and quark-photon vertex (QPV), respectively. These pieces are obtained self-consistently within the CI framework presented below. For simplicity, we have omitted the flavor indices.

\subsection{Contact Interaction: Highligths}
\label{sec:SCI}
To evaluate the elastic and transition form factors, Eqs.\,\eqref{eq:impulse} and\,\eqref{eq:IA}, we shall employ the symmetry-preserving contact interaction (CI) model described in Ref.\,\cite{Xing:2021dwe}, which incorporate components beyond the typical RL approximation\,\cite{Roberts:2010rn,Gutierrez-Guerrero:2010waf}. 

The first element to be described is the DSE equation for the quark propagator, namely the gap equation. This reads:
\begin{equation}\label{eqn:gap}
    S^{-1}(p)=i\gamma \cdot p+m+\frac{4}{3m_G^2}\int_q \gamma_\mu S(q) \gamma_\mu\,.
\end{equation}
Here $m$ is the current quark mass, and $m_G$ sets the mass scale governing the strength of the interaction. The above integral is divergent, which reflects the fact that the CI model is non-renormalizable and thus requires the implementation of a regularization scheme. We adopt the symmetry-preserving approach presented in \cite{Xing:2022jtt}, and summarized in Appendix\,\ref{ap:regu}. This procedure preserves essential symmetries, including the identities of Eq.\,\eqref{eq:wtisf}, and the gauge invariance of the $\mathcal{A}_{\mu\nu\rho}$ transition amplitude. The so-called gauge symmetry consistency relations are fulfilled as well,\,\cite{Wu:2002xa,Wu:2003dd}, and special care is taken with the $\gamma_5$ matrix, whose treatment is particularly subtle\,\cite{Korner:1991sx,Jegerlehner:2000dz,Belusca-Maito:2023wah,tHooft:1972tcz,Breitenlohner:1975hg}. Irrespective of the scheme, any consistent regularization produces the following form for the dressed quark propagator:
\begin{equation}\label{eqn:quark}
    S^{-1}(p)=i\gamma \cdot p+M\,.
\end{equation}
Evidently, the mass function $M$ turns out to be independent of the quark momentum $p$, adopting the role of a constituent quark mass.

The regularization procedure employed herein incorporates infrared and ultraviolet proper-time regulators, $\tau_{ir}:= \Lambda_{ir}^{-1}$ and $\tau_{uv}:=\Lambda_{uv}^{-1}$, respectively.  The former, $\Lambda_{ir}\approx\Lambda_{\text{QCD}}$, ensures the absense of quark production thresholds, producing a picture compatible with confinement;  $\Lambda_{uv} \approx m_p$ ($\sim$ proton mass) is irremovable and fixes the scale of all dimensionful quantities. It is well established that the presence of a finite cutoff in effective field theories can lead to only partial realization of the expected anomalies\,\cite{Alkofer:1992nh}. However, it has been suggested that higher-order corrections designed to emulate the underlying suppressed dynamics can fix this issue\,\cite{Lepage:1997cs}. Following these ideas, we employ the modified rainbow-ladder (MRL) truncation developed in Ref.\,\cite{Xing:2021dwe}. Within this framework, an anomalous magnetic moment (AMM) term is dynamically generated in the QPV. The infrarred strength of such is governed by DCSB\,\cite{Chang:2010hb,Bashir:2011dp}. Consequently, it has been shown to restore the chiral anomaly associated to the $\pi^0\to\gamma\gamma$ and $\gamma\to 3\pi$ processes\,\cite{Dang:2023ysl,Xing:2024bpj}, effectively capturing the short-distance dynamics left out by the ultraviolet cutoff. Its robustness has also been tested in the $\eta(\eta') \to \gamma \gamma$ process and the vector to pseudoscalar radiative decays\,\cite{Xu:2024frc}.

Taking all these observations into account, we express the homogeneous BS equation  for an arbitrary meson $H$ as follows:
\begin{eqnarray}\label{eq:BSE}
    \Gamma_{H}(Q)&=&-\frac{4}{3m_G^2}\int_q \gamma_\alpha S(q)\Gamma_{H}(Q)S(q-Q)\gamma_\alpha\nn
    &+&\frac{4\xi}{3m_G^2}\int_q \tilde{\Gamma}_j S(q)\Gamma_{H}(Q)S(q-Q)\tilde{\Gamma}_j\,.
\end{eqnarray}
The first line in the above equation stems from the RL truncation. The second line contains the beyond RL contribution, defined by the tensors $\tilde{\Gamma}_j=\left\{I_4,\gamma_5,\frac{i}{\sqrt{6}}\sigma_{\al\be}\right\}$, and weighted by the dimensionless parameter $\xi$. As can be noted, the meson BSA depends solely on the meson's total momentum $Q$ (with $Q^2=-m_H^2$). There is no dependence on the relative momentum between the valence quark and antiquark. This, together with the momentum-independent mass function, is an essential characteristic of the CI. 

The general tensor decomposition of the BSA is determined by the meson's  quantum numbers. For example, for the AV meson:
\begin{eqnarray}
    \label{eq:bsaAxial}
    \Gamma^{\text{A}}_\mu(Q)&=&\gamma_5\gamma^T_\mu E_{\text{A}}(Q)\,,
\end{eqnarray}
In this case, the combination of Eqs.\,\eqref{eq:BSE} and\,\eqref{eq:bsaAxial} yields ($\bar{\alpha}=1-\alpha$):
\begin{eqnarray}
    1&=&\mathcal{K}(P^2=-m_\text{A}^2)\,,\\
    \mathcal{K}(P^2)&=&-\frac{16}{3m_G^2}\int_0^1d\alpha (M^2+\alpha\bar{\alpha}P^2) I_0(M^2+\alpha\bar{\alpha}P^2)\,,\nonumber
\end{eqnarray}
from which one can identify the AV meson mass $m_\text{A}$. The resulting amplitudes must be then normalized according to the canonical condition,\,\cite{Nakanishi:1969ph}, such that:
\begin{equation}
    \frac{1}{E_\text{A}^2}=-\frac{9m_G^2}{2} \frac{d \mathcal{K}(P^2)}{dP^2}|_{P^2=-m_\text{A}^2}
\end{equation}

Within the present scheme, the QPV is obtained from the corresponding inhomogeneous BS equation:
\begin{eqnarray}\label{eq:QPV}
    \Gamma_{\mu}(Q)&=&\gamma_\mu-\frac{4}{3m_G^2}\int_q \gamma_\alpha S(q)\Gamma_{\mu}(Q)S(q-Q)\gamma_\alpha\nn
    &+&\frac{4\xi}{3m_G^2}\int_q \tilde{\Gamma}_j S(q)\Gamma_{\mu}(Q)S(q-Q)\tilde{\Gamma}_j\,.
\end{eqnarray}
 The structure acquired by the QPV is:
\begin{equation}
\label{eq:QPVstru}
    \Gamma_{\mu}(Q)=\gamma^L_\mu \hat{f}_L(Q^2)+\gamma^T_\mu \hat{f}_T(Q^2)+\frac{\sigma_{\mu\nu}Q_\nu}{M} \hat{f}_A(Q^2)\,,
\end{equation}
where $\gamma_\mu^T=\gamma_\mu-\frac{\slashed{Q}Q_\mu}{Q^2}$, $\gamma_\mu^L=\gamma_\mu-\gamma_\mu^T$. Plugging it into Eq.\,\eqref{eq:QPV}, it can be readily shown that $\hat{f}_L(Q^2) = 1$. The remaining part of the vertex has a similar structure to that of the vector meson BSA, 
therefore both $\hat{f}_{T,A}(Q^2)$ exhibit a vector meson pole in the timelike axis ($Q^2=-m_{\text{V}}^2$). Moreover,  the AMM-related term $\hat{f}_A(Q^2)$ is zero when $\xi = 0$, reflecting that its existence emerges from contributions beyond RL. Additional chracteristics of the QPV dressing functions are examined in Ref.\,\cite{Xing:2021dwe,Xu:2024frc}.

\section{Results and Discussion}
\label{sec:Results}

\subsection{Parameters and Mass Spectrum}

Being focused on the lightest \text{AV} mesons, namely $a_1,f_1,f_1'$, we adopt CI model parameters typical for the light-sector\,\cite{Xing:2022jtt,Xing:2021dwe}. First, the parameters defining the CI model are fixed as:
\begin{equation}
    \label{eq:modparams1}
    m_G=0.132\,\text{GeV}\,,\Lambda_{ir}=0.240\,\text{GeV}\,,\,\Lambda_{uv}=0.905\,\text{GeV}\,.
\end{equation}
In addition, $\xi=0.151$ is chosen to ensure the fulfillment of the chiral anomaly\,\cite{Dang:2023ysl,Xing:2024bpj}. Note the variation of $\xi$ only affects the vector channels, e.g. the vector meson masses. Assuming isospin symmetry $m_{u}=m_d$, and seeking to reproduce the pion and kaon masses, we set $m_{u/d}=0.007$ \text{GeV} and $m_s=0.170$ \text{GeV}. This setup yields the constituent quark and meson masses collected in Table\,\ref{tab:Masses}.

\begin{table}[h!]
\centering
\caption{Quark and meson masses (in GeV)}
\begin{tabular}{c c | c c c c c c}
\hline
$m_u$ & $m_s$ & $M_u$ & $M_s$ & $m_\pi$ & $m_K$ & $m_\rho$ & $m_\phi$\\
\hline
$0.007$ & $0.170$ & $0.368$ & $0.533$ & $0.140$ & $0.499$ & $0.879$ & $1.058$ \\
\hline
\end{tabular}
\label{tab:Masses}
\end{table}

It is important to highlight that the beyond RL contribution favors a better description of the vector channels, and so the QPV, but it does not contribute at all to the axial-vector case \cite{Xing:2021dwe}. Accordingly, \text{AV} mesons are described solely by the RL truncation, which is known to fail because it omits spin-orbit repulsion effects expected to be relevant in states with non-negligible components of orbital angular momentum\,\cite{Lu:2017cln,Yin:2021uom,Gutierrez-Guerrero:2021rsx,Qin:2020jig,Chang:2009zb}. Among other issues, this results in  underestimated \text{AV} meson masses. A typical solution within the CI framework is to multiply the RL kernel by a dimensionless coupling $g_{\text{SO}}^2\leq1$ in these channels\,\cite{Lu:2017cln,Yin:2021uom,Gutierrez-Guerrero:2021rsx,Chen:2012qr}. A smaller $g_{\text{SO}}$ translates into a larger mass, while the original result is clearly recovered with $g_{\text{SO}}=1$. 

\begin{figure}[t]
\centerline{%
\begin{tabular}{c}
\includegraphics[width=0.5\textwidth]{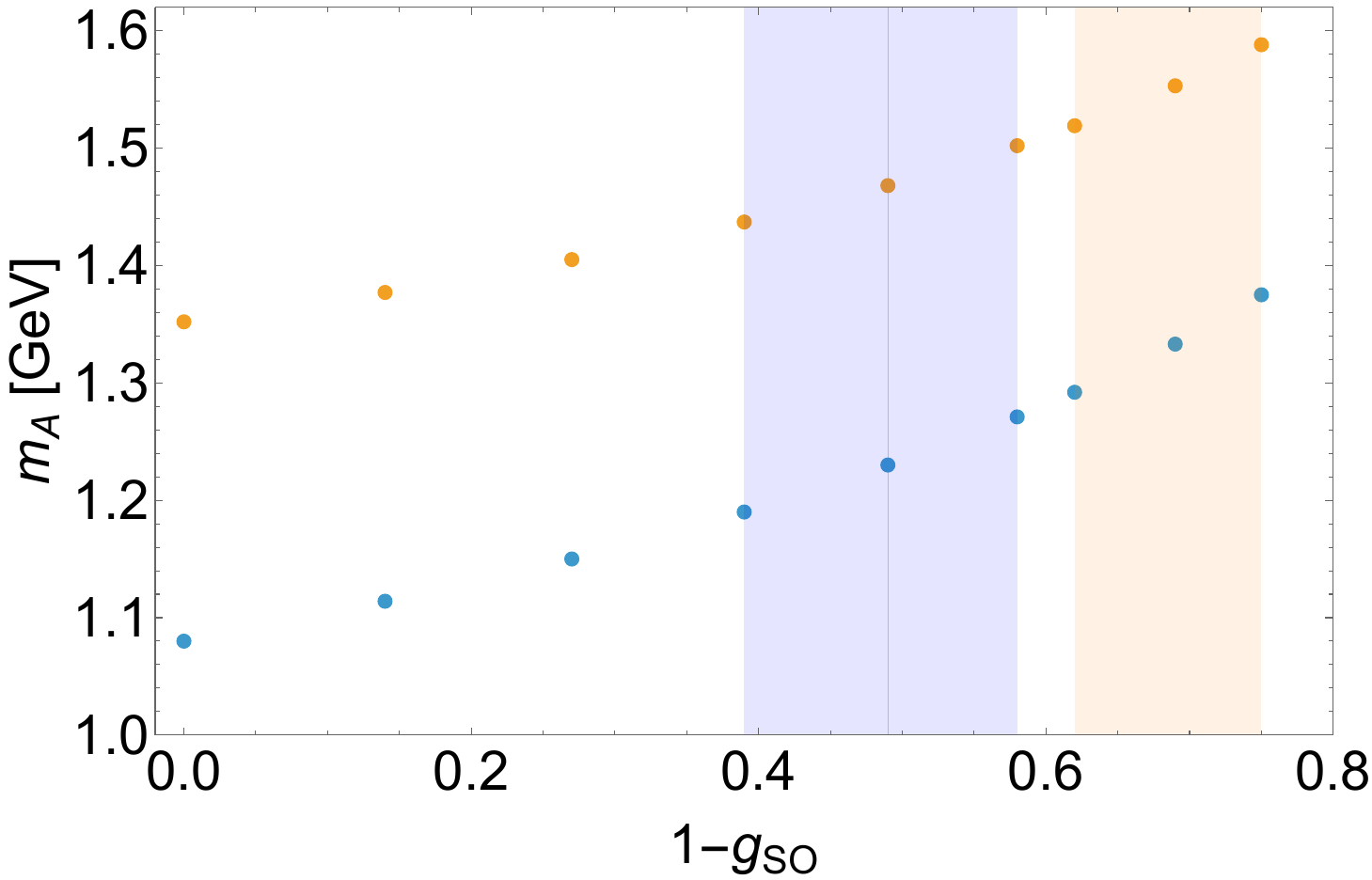} 
\end{tabular}}
\caption{Computed $m_{a_1}$ (blue) and $m_{f_1'}$ (orange) masses as a function of the spin-orbit repulsion parameter $g_{\text{SO}}$.  The blue and orange bands point out to $m_{a_1}=1.230(40)\,\text{GeV}$ and $m_{a_1}=1.333(41)\,\text{GeV}$, corresponding to the empirical values of $m_{a_1}$ and $m_{a_1}-m_\rho$, respectively\,\cite{ParticleDataGroup:2024cfk}.  }
\label{fig:MassDependence}     
\end{figure}

The impact of spin-orbit coupling on the axial meson masses is illustrated in Fig.\,\ref{fig:MassDependence}, with the relevant numerical values summarized in Table\,\ref{tab:MassesAV}. We consider three cases: $g_{\text{SO}}$ fixed to either produce the experimental $m_{a_1}-m_\rho$ mass-splitting (Case-I) or the empirical value $m_{a_1}$ (Case-II),\,\cite{ParticleDataGroup:2024cfk}; and $1-g_{\text{SO}}=0$, corresponding to the omission of spin-orbit repulsion effects (Case-O). These three cases will be used later for the analysis of the form factors and their contribution to $a_\mu$. Note, due to isospin symmetry ($m_u=m_d$) and no mixing between the $f_1$ and $f_1'$ sates, the $a_1$ and $f_1$ mesons become mass degenerarate while the $f_1'$ is described as a purely $s\bar{s}$ state.

\begin{table}[h!]
\centering
\caption{Axial-vector meson masses (in GeV), and the spin-orbit coupling $g_{\text{SO}}$. The underlined quantities indicate the experimental values used to determine $g_{\text{SO}}$\,\cite{ParticleDataGroup:2024cfk}. }
\begin{tabular}{l | c c c c}
\hline
 & $m_{a_1}$ & $m_{f_1'}$ & $m_{a_1}-m_\rho$ & $1-g_{\text{SO}}$ \\
\hline
Case-I & $1.333(41)$ & $1.553(35)$ & \underline{$0.452(41)$} & $0.69(7)$\\
Case-II & \underline{$1.230(41)$} & $1.469(31)$ & $0.351(40)$ & $0.49(10)$\\
Case-O & $1.080$ & $1.352$ & $0.200$ & $0$\\
\hline
\end{tabular}
\label{tab:MassesAV}
\end{table}

\subsection{Elastic form factors}
\begin{figure}[t]
\centerline{%
\begin{tabular}{c}
\includegraphics[width=0.5\textwidth]{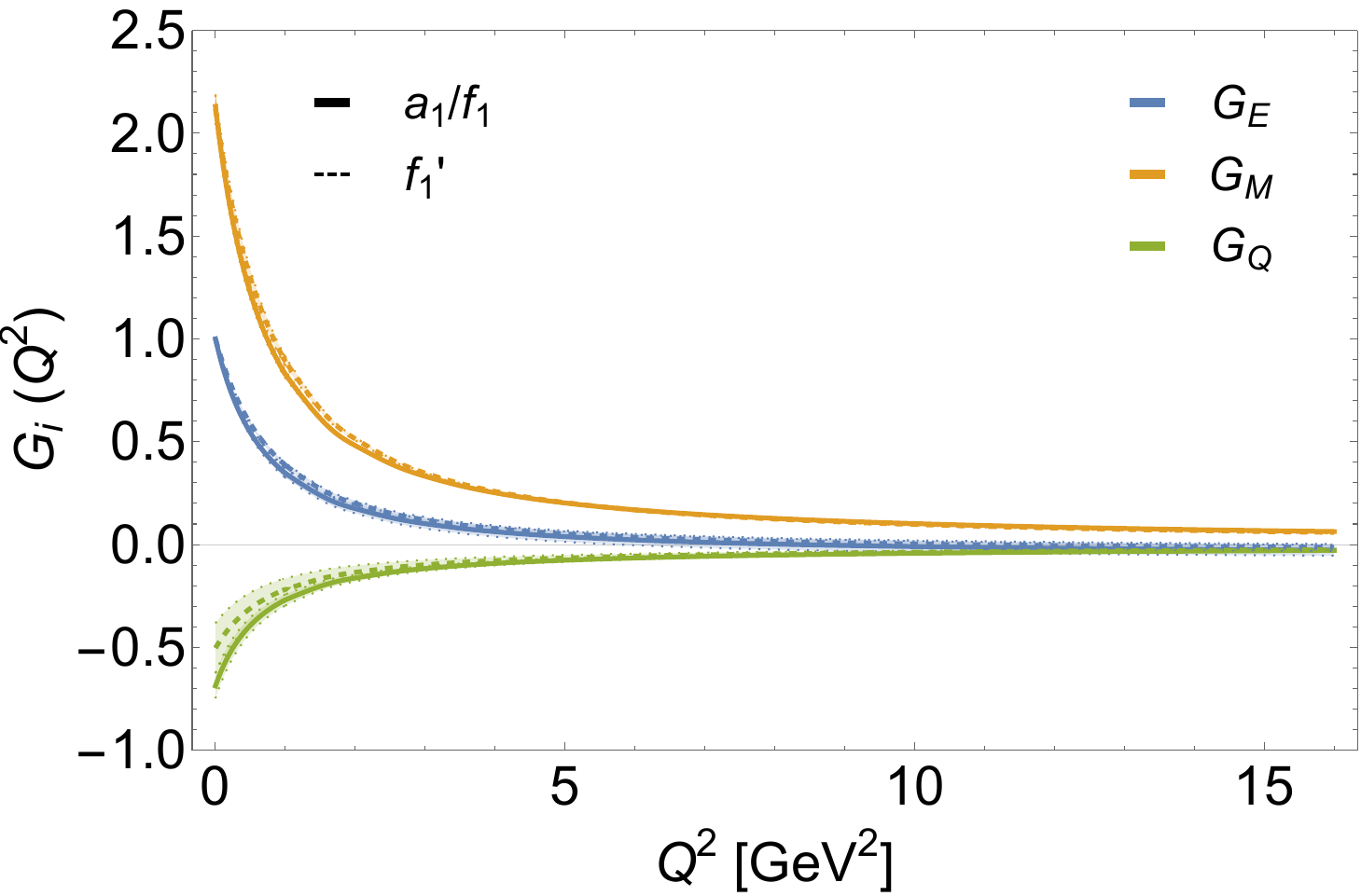} 
\end{tabular}}
\caption{Axial-vector meson EFFs. The band reflects the variation of $m_{a_1}$ and $m_{f_1'}$ within the whole range consider here (see Table\,\ref{tab:MassesAV}).}
\label{fig:EFFs}     
\end{figure}

Let us consider the positively charged state $a_1^+$ and, for the purposes of the discussion, the $f_1$ and $f_1'$ mesons are likewise treated as if they were positively charged. The same applies to the $\rho^+$ meson and its $s\bar{s}$ counterpart, the $\phi$ meson. Within the present scheme, $a_1$ and $f_1$ are thus identical and the resulting EFFs are the same. These, together with the $f_1'$ case are depicted in Fig.\,\ref{fig:EFFs}. It is evident that, within the range of variation for the AV meson masses in Table\,\ref{tab:MassesAV}, the EFFs exhibit little sensitivity.

The $Q^2 \to 0$ limit of the EFFs, Eq.\,\eqref{eq:GEMQ}, defines the charge, magnetic and quadrupole moments:
\begin{eqnarray}
\label{eq:Static1}
G_E(0)=1\,,\,G_M(0)=\mu\,,\,G_{Q}(0)=\mathcal{Q}\,.
\end{eqnarray}
The produced values are collected in Table\,\ref{tab:Static}, in addition to the corresponding ones for the vector mesons $\rho$ and $\phi$. The magnetic moments are essentially the same for the AV mesons under examination, $\mu\approx2.13$, and somewhat smaller than those of their vector meson counterparts. The $f_1'$ quadrupole moment exhibits a $\sim25\%$ reduction in magnitude with respect to $a_1$. This pattern in repeated in the vector meson case and, in fact, we find $\mathcal{Q}_{a_1}\approx \mathcal{Q}_\rho$ and $\mathcal{Q}_{f_1'}\approx \mathcal{Q}_\phi$.

It is also revealed by Fig.\,\ref{fig:EFFs} that, despite being somewhat compatible within errors, the $f_1'$ form factors exhibit a slightly less pronounced large-$Q^2$ falloff. This can be quantified through the corresponding radii, defined as:
\begin{equation}
\label{eq:radii}
    r_i^2=-6\frac{1}{G_{i}(0)}\frac{\partial G_{i}(Q^2)}{\partial Q^2}\Bigg|_{Q^2\to0}\,.
\end{equation}
The results of this evaluation are presented in Table,\ref{tab:Static}. Unsurprisingly, for all radii shown, we find $r^{f_1'}_i<r_i^{a_1}$. Moreover, as with the magnetic and quadrupole moments, $r_i^{a_1}\approx r_i^{\rho}$ and $r_i^{f_1'}\approx r_i^\phi$. Although the vector-meson case is not shown explicitly, all these quantities reflect the similarity between the EFFs of vector and AV mesons in the low-$Q^2$ region.

Turning our attention to $G_E(Q^2)$, we observe a sign flip near $Q^2 \approx 11$ and $13~\text{GeV}^2$ for the $a_1$ and $f_1'$ mesons, respectively. This zero-crossing reflects the predicted large-$Q^2$ behavior of spin-1 mesons,\,\cite{Haberzettl:2019qpa,Brodsky:1992px}, and arises from the destructive interference between the $G_{1,2,3}$ for factors defining $G_E$, Eq.\,\eqref{eq:GEMQ1}. However, for the $\rho$ and $\phi$ mesons, the zero-crossing occurs much more rapidly: at around $Q^2 = 3.5$ and $5~\text{GeV}^2$, respectively. The specific values are listed in Table\,\ref{tab:Static}. When expressing this sign change in terms of $x_c=Q_c^2/m_H^2$, such that $G_E^H(Q_c^2)=0$, the value of $x_c$ is smaller for the heavier system. This pattern is consistent with the findings of Ref.\,\cite{Xu:2019ilh}, which employs a more sophisticated numerical treatment for the study of vector mesons. Thus, although the EFFs of vector and axial-vector mesons are similar in the low-virtuality domain, differences emerge as $Q^2$ increases. This would result, for instance, in into markedly different charge distributions,\,\cite{Xu:2025sxw}.

\begin{table}[h!]
\centering
\caption{Static properties of axial-vector and vector mesons. Herein $x_c=Q_c^2/m_H^2$, defined by the zero-crossing $G_E^H(Q_c^2)=0$. The error in the AV case stems from the variation on the corresponding masses (see Table\,\ref{tab:MassesAV}).}
\begin{tabular}{l | c c c c c c c}
\hline
 & $\mu$ & $\mathcal{Q}$ & $r_E$/fm & $r_M$/fm & $r_{Q}$/fm & $x_c$ \\
\hline
$a_1$ & 2.12(5) & -0.69(6) & -0.59(2) & 0.57(1) & 0.58(1) & 6.55(1.61)\\
$f_1'$ & 2.13(6) & -0.50(12) & -0.53(1) & 0.52(1) & 0.52(1) & 5.61(1.43)\\
$\rho$ & 2.22 & -0.63 & -0.61 & 0.55 & 0.56 & 4.57\\
$\phi$ & 2.26 & -0.53 & -0.51 & 0.47 & 0.48 & 4.38\\
\hline
\end{tabular}
\label{tab:Static}
\end{table}

\subsection{Transition form factors}
As in the case of EFFs, our present framework produces identical $a_1$ and $f_1$ TFFs, differing solely by the flavor-color factors $N_{fc}$ given in Eq.\,\eqref{eq:flavorNorm}. The $f_1'$ remains a purely $s\bar{s}$ state. Thus, considering first Case-I, where the experimental value of $m_{a_1}-m_\rho$ is reproduced, in the limit $Q^2,Q'^2\to 0$ we obtain:
\begin{equation}
\label{eq:F0I}
    F_{1,2,3}^{\text{A}}(0,0)=\,0.554^{+0.018}_{-0.012}\,,\; 0.280^{+0.009}_{-0.006}\,,\; -0.006^{+0.001}_{-0.001}\,.
\end{equation}
Each $F_i^\text{A}$ arises from adding the individual contriutions of $\{a_1,f_1,f_1'\}$ with their corresponding $N_{fc}$ weights. When we consider Case-II instead, with the empirical value for $m_{a_1}$, our exploration leads us to:
\begin{equation}
\label{eq:F0II}
    F_{1,2,3}^{\text{A}}(0,0)=\,0.532^{+0.005}_{-0.004}\,,\; 0.263^{+0.008}_{-0.006}\,,\; -0.004^{+0.001}_{-0.001}\,.
\end{equation}
The asymmetric errors in both Case-I and Case-II account for the experimental uncertainty associated to $m_{a_1}-m_\rho$ and $m_{a_1}$, respectively. Finally, the omission of spin-orbit repulsion in the interaction kernels, characterized by $m_{a_1}=1.080$ GeV, yields:
\begin{equation}
\label{eq:F0O}
   F_{1,2,3}^{\text{A}}(0,0)=\,0.518\,,0.230\,,-0.002\,.
\end{equation}
It is worth noting that the magnitude of all $F_i^{\text{A}}$ exhibit a mild increasing trend with $m_{a_1}$ and, in fact, the TFFs in the whole range are just mildly affected by the variation of the AV meson mass. This is a common feature with the corresponding EFFs. The resulting $F_3^{\text{A}}(0,0)$ are practically negligible, despite the fact that it exhibits the largest relative variation among the different setups.  By contrast, the variation range is below $20\%$ for $F_2^{\text{A}}(0,0)$, and under $10\%$ for $F_1^{\text{A}}(0,0)$. Therefore, in the remainder of the discussion we will consider an average that accounts for the entire range of $m_{a_1}$ values being considered (see Fig.\,\ref{fig:MassDependence} and Table\,\ref{tab:MassesAV}). This leads us to the following:
\begin{equation}
\label{eq:Fave}
F_{1,2,3}^{\text{A}}(0,0)=\,0.545(27)\,,0.259(30)\,,-0.005(2)\,.
\end{equation}
According to the analysis of Ref.\,\cite{Eichmann:2024glq}, which makes use of data-driven TFF parametrizations from Refs.\,\cite{Hoferichter:2023tgp,Zanke:2021wiq}, one expects the following constraints:
\begin{equation}
\label{eq:ConstFFs}
F_{1,2,3}^{\text{A}}(0,0) = \begin{bmatrix}
2.45(35) \\[6pt]
1.96(42)
\end{bmatrix},\,\begin{bmatrix}
2.18(1.17) \\[6pt]
0.42(1.49)
\end{bmatrix},\,\begin{bmatrix}
-0.74(84) \\[6pt]
-0.33(84)
\end{bmatrix}.
\end{equation}
Clearly, our expectations for $F_{1,2,3}^{\text{A}}(0,0)$ are underestimated, with the latter being particularly notable. It can be thus anticipated that the resulting contribution to $a_\mu$ will be smaller than expected. Increasing $\xi=0.151\to0.38$ in the BS kernel, Eq.\,\eqref{eq:BSE}, allows one to reproduce the physical $\rho$-meson mass without affecting that of the AV meson. This modification enhances $F_i(0,0)$ by approximately $50\%$, $80\%$, and $600\%$ for $i=1,2,3$, respectively. This would certainly bring our expectations closer to the constraints in Eq.\,\eqref{eq:ConstFFs}. On the other hand, in a RL-like setup ($\xi=0$, $m_\rho=0.929$ GeV), the situation would be much worse.  We choose to retain our preferred value of $\xi=0.151$, as its determination is dictated by the chiral anomaly,\,\cite{Dang:2023ysl,Xing:2024bpj}. Another contributing factor for this shortcoming arises from the CI model itself, where axial-vector mesons are described by a single BSA and vector mesons by two. This approach oversimplifies the rich tensor structure that characterizes the BSAs of these systems,\cite{Hilger:2015ora}. In an ideal scenario, one would use a truncation that captures both symmetry requirements (e.g. the chiral anomaly) and the properties of the vector and axial-vector mesons, guaranteeing that the AV TFFs fulfill the low-energy constraints.

\begin{figure}[t]
\centerline{%
\begin{tabular}{c}
\includegraphics[width=0.5\textwidth]{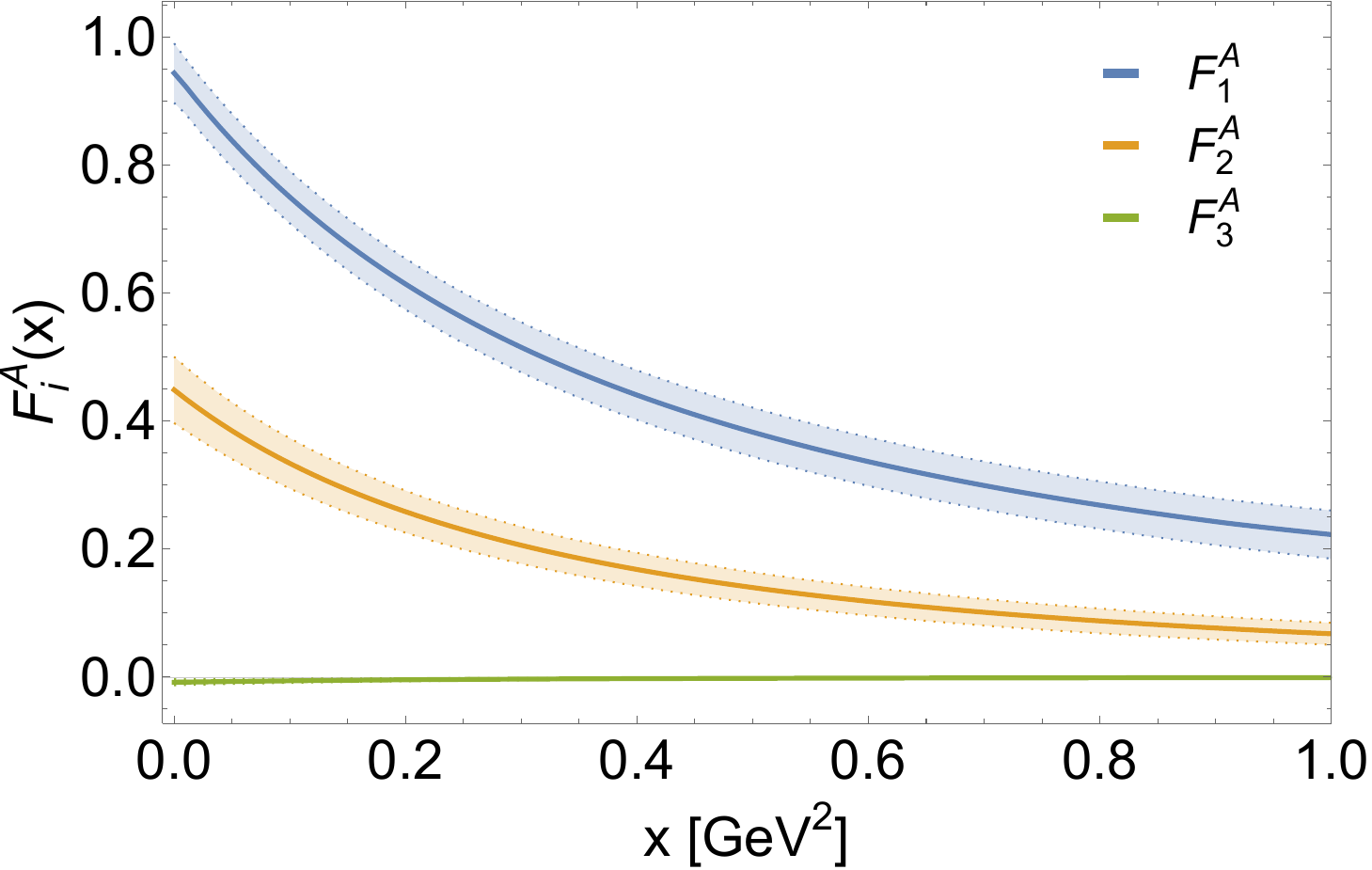} 
\end{tabular}}
\caption{Axial-vector meson to two-photon TFFs, as a function of $x=(Q^2+Q'^2)/2$. The band reflects both the variation on $m_{a_1}$ and the kinematic range from the symmetric ($Q^2=Q'^2$) to the asymmetric ($Q^2=0$ or $Q'^2=0$) limits.}
\label{fig:TFFs}     
\end{figure}

The photon virtuality dependence of the TFFs is illustrated in Fig.\,\ref{fig:TFFs}, in the symmetric limit $Q^2=Q'^2$. The $|F_1^{\text{A}}| > |F_2^{\text{A}}| > |F_3^{\text{A}}|$ hierarchy is readily apparent. This ordering persists over a broad range, with all TFFs decreasing in magnitude as $Q^2$ increase. Notably, this falloff is much sharper than that seen in the CI description of the pion elastic and transition form factors\,\cite{Gutierrez-Guerrero:2010waf,Roberts:2010rn,Dang:2023ysl}. To quantify this rate, one may define transition radii $r_i^{\text{A}\gamma}$ in analogy to Eq.\,\eqref{eq:radii}, which yield:
\begin{equation}
    \label{eq:Radii}
    r_{1,2,3}^{\text{A}\gamma}=0.76(1),\,0.86(1),\,0.95(1)\,\text{fm}\,.
\end{equation}
The AV TFFs display not only a noticeable decrease with momentum transfer but also the correct asymptotic power-law behavior. On the other hand, the CI model does not reproduce certain aspects of perturbation theory, including the anomalous dimensions of the form factors. This limitation is expected to have little impact, since the kinematic region of the TFFs relevant to $a_\mu$ is restricted to $Q^2,Q'^2 \lesssim 5\,\text{GeV}^2$. In any case, to analyze their contributions to $a_\mu$, we adopt the QCD-driven parametrization of Ref.\,\cite{Eichmann:2024glq}. The explicit form of such and its corresponding interpolation parameters are discussed in Appendix\,\ref{ap:Params}.

Finally, it is worth pointing out that the detailed breakdown of the TFFs into $a_1, f_1,$ and $f_1'$ contributions is omitted here, as $a_1$ and $f_1$ possess identical form factors and the $f_1'$ ones are quite similar in shape. For reference, the $f_1'$ contribution amounts to about $10$–$25\%$ of the whole. Alternatively, assuming that the TFFs of all these mesons are identical, Eq.\,\eqref{eq:flavorNorm} indicates that $f_1'$ would account for $\sim18\%$ of the total.

\subsection{Hadronic light-by-light contributions}
For the computation of the HLbL contribution, we consider the master formula presented in\,\cite{Colangelo:2017fiz}:
\begin{equation}
\label{eq:Master}
\alpha_{\mu}^{\text{HLbL}} = \frac{\alpha^3_{em}}{432 \pi^2} \int_{\Omega}  \sum_i^{12} T_i(Q_1,Q_2,Q_3) \bar{\Pi}_i (Q_1,Q_2,Q_3)\,,
\end{equation}
where $\alpha_{em}$ is the QED coupling constant, and
\begin{equation}
    \int_{\Omega}:= \int_{0}^{\infty} d\Sigma~\Sigma^3 \int_0^1 dr~r\sqrt{1-r^2} \int_0^{2\pi}d\phi\;.
\end{equation}
The integration variables $\{\Sigma,r,\phi\}$ are related with the (Euclidean) photon virtualities, $Q_{1,2,3}^2$, as follows:
\begin{eqnarray}
    Q_1^2&=&\frac{\Sigma}{3}\left(1-\frac{r}{2}\cos \phi - \frac{r}{2}\sqrt{3}\sin \phi \right)\,,\nonumber\\
    Q_2^2&=&\frac{\Sigma}{3}\left(1-\frac{r}{2}\cos \phi + \frac{r}{2}\sqrt{3}\sin \phi \right)\,,\nonumber\\
    Q_3^2&=&\frac{\Sigma_3}{3}(1+r\cos\phi)\,.
\end{eqnarray}
The kernel functions $T_i$ entering Eq.\,\eqref{eq:Master} are well known. These are of purely kinematic nature and are provided in the appendix of Ref.\,\cite{Colangelo:2017fiz}. The dynamical component, specific to each type of HLbL contribution, is encoded in the twelve scalar functions $\bar{\Pi}_i$. As discussed in Ref.\,\cite{Hoferichter:2024fsj}, these can be expressed via six representative functions $\hat{\Pi}_i$. The case of the AV meson contribution is presented in\,\cite{Hoferichter:2024fsj}, in terms of form factors $\mathcal{F}_i$. When translated into the form factors $F_i$ defined in Eqs.\,\eqref{eq:Fis}, these read:
\begin{eqnarray}
    \mathcal{F}_1(Q^2,Q'^2)&=&-\frac{Q^2-Q'^2}{2m_\text{A}^2}F_3\,,\nonumber\\
    \mathcal{F}_2(Q^2,Q'^2)&=&\frac{1}{2}\left(F_1-\frac{Q^2-Q'^2}{2m_\text{A}^2}F_2\right)\,,\nonumber\\
    \mathcal{F}_3(Q^2,Q'^2)&=&-\frac{1}{2}\left(F_1+\frac{Q^2-Q'^2}{2m_\text{A}^2}F_2\right)\,.
\end{eqnarray}
Numerical integrations are performed using the CUBA library\,\cite{Hahn:2004fe}. To calibrate and benchmark our approach, we compute the pion pole and box contributions, where the associated form factors are derived consistently within the CI formalism\,\cite{Dang:2023ysl}. By employing Padé-type parametrizations for each form factors\,\cite{Masjuan:2017tvw}, both considered at different orders, we obtain:
\begin{equation}
    \label{eq:boxci}
    a_\mu^{\pi-\text{box}}= -21.2(3)\times 10^{-11}\,,\,a_\mu^{\pi-\text{pole}}= 94.1(4)\times 10^{-11}\,.
\end{equation}
The uncertainty combines both the parametrization and numerical errors. As can be noted, the produced values are about $30-50\%$ larger than typical estimates, including DSE evaluations,\,\cite{Raya:2019dnh,Eichmann:2019tjk,Miramontes:2021exi,Miramontes:2024fgo,Eichmann:2019bqf}, and the 2025 White Paper (WP2) consensus\,\cite{Aliberti:2025beg}. For instance, $a_\mu^{\pi-\text{box}}[\text{WP2}]=-15.9(2)\times 10^{-11}$. This outcome is not unexpected given the fact that the CI not only produces hard form factors but, in these cases, also drives their asymptotic behavior toward a constant\,\cite{Gutierrez-Guerrero:2010waf,Roberts:2010rn}, failing to fulfill the asymptotic expectations\,\cite{Lepage:1980fj}. Furthermore, the CI model delivers a flat-top valence-quark distribution function,\,\cite{Gutierrez-Guerrero:2010waf}, corresponding to one of the physical bounds identified for this system,\,\cite{Lu:2023yna}. The results in Eq.\,\eqref{eq:boxci} should therefore be interpreted as limiting cases rather than precise estimates for these quantities.

\begin{figure}[t]
\centerline{%
\begin{tabular}{c}
\includegraphics[width=0.5\textwidth]{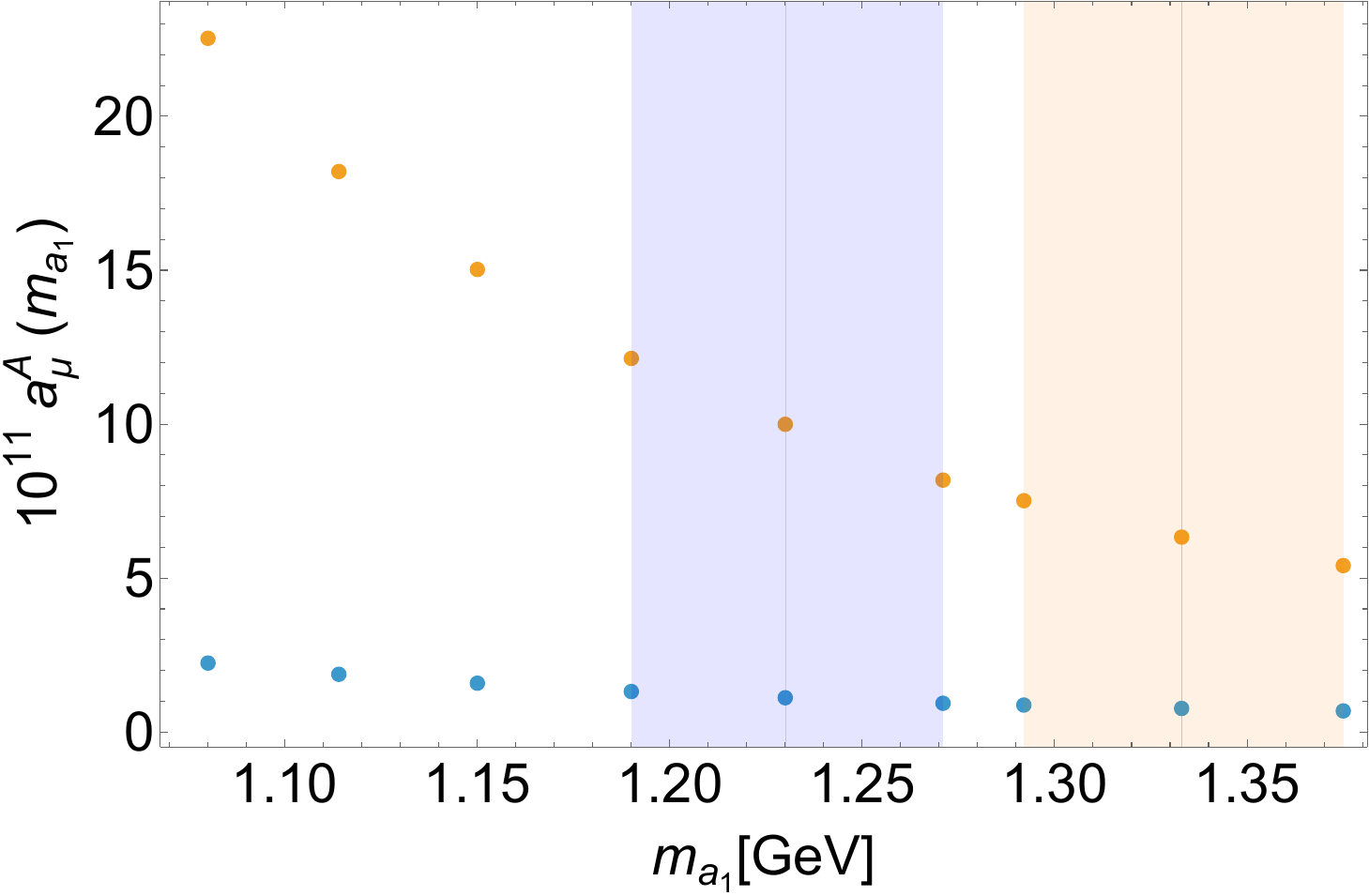} 
\end{tabular}}
\caption{AV meson mass dependence on $a_\mu^{\text{A}}$, obtained form the original (blue) and rescaled (orange) TFFs. The blue and orange bands point out to $m_{a_1}=1.230(40)\,\text{GeV}$ and $m_{a_1}=1.333(41)\,\text{GeV}$, corresponding to the empirical values of $m_{a_1}$ and $m_{a_1}-m_\rho$, respectively\,\cite{ParticleDataGroup:2024cfk}.  }
\label{fig:MassAmu}     
\end{figure}

Turning our attention to the AV meson case, and considering solely the lowest-lying multiplet, $\{ a_1, f_1, f_1'\}$, we arrive at the following:
\begin{subequations} \label{eq:amuUn}
\begin{align}
a_\mu^{\text{A}}[\text{Case-I}]=&\,0.77_{-0.08}^{+0.11}\times 10^{-11}\,, \label{eq:amuUn1} \\
a_\mu^{\text{A}}[\text{Case-II}]=&\,1.14_{-0.18}^{+0.20}\times 10^{-11}\,,\label{eq:amuUn2} \\
a_\mu^{\text{A}}[\text{Case-O}]=&\,2.24\times 10^{-11}\,.\label{eq:amuUn3}
\end{align}
\end{subequations}
Here, the asymmetric error accounts for the variation of $g_{\text{SO}}$ to produce the experimental mass splitting $m_{a_1}-m_\rho$ in Case-I, and the empirical value $m_{a_1}$ in Case-II. Recall Case-O corresponds to neglecting the spin-orbit coupling ($g_{\text{SO}}=1$), hence producing a smaller AV mass, $m_{a_2}=1.08\,\text{GeV}$ (see Table\,\ref{tab:MassesAV}). The numerical and parametrization error estimates are minimal and shall be omitted in the discussion. These results are consistent with the 2020 White Paper (WP1)\,\cite{Aoyama:2020ynm}, $a_\mu^\text{A}[\text{WP1}] = 6(6)\times10^{-11}$, but largely disfavored by more recent analysis. This includes the WP2\,\cite{Aliberti:2025beg}, the discussion in Ref.\,\cite{Masjuan:2020jsf}, and the DSE evaluation from\,\cite{Eichmann:2024glq}. All of them tend to favor a larger value:
\begin{subequations} \label{eq:amuVarious}
\begin{align}
a_\mu^{\text{A}}[\text{WP2}]=&\,15.8(1.6)\times 10^{-11}\,, \label{eq:amuVarious1} \\
a_\mu^{\text{A}}[\text{DSE}]=&\,17.4(4.0)(4.0)(0.5)\times 10^{-11}\,,\label{eq:amuVarious2} \\
 a_\mu^{\text{A}}[\text{Ref.\,\cite{Masjuan:2020jsf}}]=&\,16.0_{-4.5}^{+5.1}\times 10^{-11}\,.\label{eq:amuVarious3}
\end{align}
\end{subequations}
This places our estimates below recent trends. This underestimation is evidently driven by the value of the form factors at $Q^2, Q'^2 \to 0$, Eq.\,\eqref{eq:Fave}. To address this issue, and in line with Ref.\,\cite{Eichmann:2024glq}, we rescale the TFFs so as to enforce:
\begin{equation}
\label{eq:Decay}
    |F_1^{\text{A}}(0,0)|^2=\frac{48 \Gamma_{\gamma\gamma}}{\pi \alpha_{em}^2 m_\text{A}}\,,
\end{equation}
with $\Gamma_{\gamma\gamma}=3.5(6)(5)\,\text{keV}$ for $f_1(1285)$\,\cite{L3:2001cyf}. Accounting only for the central value of the decay width, we find:
\begin{subequations} \label{eq:amuUnb}
\begin{align}
a_\mu^{\text{A}}[\text{Case-I}]=&\,6.33_{-0.92}^{+1.18}\times 10^{-11}\,, \label{eq:amuUn1b} \\
a_\mu^{\text{A}}[\text{Case-II}]=&\,9.99_{-1.81}^{+2.14}\times 10^{-11}\,,\label{eq:amuUn2b} \\
 a_\mu^{\text{A}}[\text{Case-O}]=&\,22.53\times 10^{-11}\,.\label{eq:amuUn3b}
\end{align}
\end{subequations}
It is evident that these estimates are closer to those in Eq.\,\eqref{eq:amuVarious}. However, as illustrated in Fig.\,\ref{fig:MassAmu}, $a_\mu^{\text{A}}$ is in fact highly sensitive to the AV meson mass. Since the variation in the magnitude of the form factors with $m_{a_1}$ is relatively modest, Eq.\,\eqref{eq:Fave}, the $1/m_{\text{A}}$ factors appearing in the HLbL master formula are primarily responsible for this behavior. To mitigate this sensitivity, we compute $a_\mu^{\text{A}}$ for a range of $m_{a_1}$ and consider the expectation value:
\begin{equation}
    \langle a_\mu^{\text{A}}\rangle := \frac{1}{m_1-m_0}\int_{m_0}^{m_1}dm_{a_1}\, a_\mu^{\text{A}}(m_{a_1})\,,
\end{equation}
with $m_0=1.080$ GeV and $m_1=1.374$ GeV (the smallest and largest $m_{a_1}$ being considered). Defining the standard deviation in the usual way, the final result is:
\begin{equation}
    a_\mu^{\text{A}}=11.30(4.71)\times10^{-11}\,.
\end{equation}
With a somewhat conservative uncertainty, this estimate is consistent with current trends. Nevertheless, it stems from TFFs with an underestimated normalization that have been rescaled to reproduce the decay width in Eq.\,\eqref{eq:Decay}. The shortcoming is naturally attributable to the limitations of the CI model (e.g. the oversimplified description of the meson BSAs). However, the most fundamental cause of the problem is the inadequacy of the interaction kernels in capturing the properties of AV mesons. As pointed out before, such kernels should also comply with symmetry and phenomenological requirements\,\cite{Chang:2009zb,Qin:2020jig}.

\section{Summary and Scope}
We computed the AV meson elastic and two photon TFFs, and the contributions of the latter to the HLbL piece of $a_\mu$. The form factors are defined in a symmetry-preserving basis and obtained via the impulse approximation, with each element therein derived within a CI framework of the DS and BS equations. By extending the usual RL truncation, this approach improves the treatment of anomalous processes (e.g., those involving the chiral anomaly) and provides a more accurate description of vector channels. This results in vector meson masses closer to their empirical value, an improved description of the QPV, and a more realistic $Q^2$ dependence of the TFFs. For the AV channels, the present approach is supplemented by an effective coupling that allows emulating the spin-orbit repulsion effects neglected in the interaction kernels. The variation of this coupling, $g_{\text{SO}}$, translates into the variation in the mass of the AV meson. Consequently, the dependence of the form factors and $a_\mu$ on the AV meson mass is also studied.

In the present context, the $a_1$ and $f_1$ mesons are mass-degenerate, while the $f_1'$ is a purely $s\bar{s}$ state. Consequently, the form factors associated with the $a_1$ and $f_1$ are identical, up to differences in the color-flavor factors in the case of the TFFs. The EFFs show little sensitivity to variations in the AV meson mass, such that the form factors associated with the $f_1'$ decrease only slightly more slowly than those of the $a_1$. In the $Q^2 \to 0$ region, which defines the corresponding radii as well as the magnetic and quadrupole moments, the EFFs of the AV mesons are, within uncertainties, fully consistent with those of their associated vector mesons, i.e. their parity partners. Nevertheless, as the photon virtuality grows, differences emerge. A notable outcome is the position of the zero-crossing of the form factor $G_E(Q^2)$, which is located in the range $Q^2\approx 3.5-5$ GeV$^2$ for the light vector mesons, but much further in the AV mesons case (above 10 GeV$^2$). This could result in a noticeably different charge distribution. These features will be explored in more detail in a future work.

A first approach to the TFFs reveal that their magnitude are underestimated. In particular, in the soft limit $Q^2,Q'^2\to0$, this effect appears moderately for $F_2$, more clearly for $F_1$, and most strongly for $F_3$, which is negative and practically negligible over the entire range of photon virtualities. These patterns will eventually translate into a rather small value of $a_\mu^\text{A}$. On the other hand, note that the $\rho$-meson mass predicted by our scheme is about $m_\rho = 0.879$ GeV. Bringing it closer to the empirical value by increasing the parameter $\xi$, which characterizes the beyond-RL contribution in the interaction kernels, significantly improves the TFF normalization. The RL limit ($\xi = 0$) yields the opposite result, as in that case $m_\rho=0.929$ GeV adopts its largest value. Given the fact that $\xi$ is set by the chiral anomaly, we keep it fixed in our analysis. Adding to this drawback, there is also the fact that, within the present CI model, the vector and axial-vector mesons are described by two and one BSAs, respectively. This is an evident oversimplication of their otherwise rich structures. Altogether, the need for interaction kernels that are robust enough to satisfy both symmetry constraints and phenomenological consistency is evident. One positive aspect is the $Q^2$ evolution of the TFFs. Contrary to what is typically observed for the CI model, these show a more visible drop, with larger transition radii and a correct power-law fall-off in the asymptotic limits.

Regarding $a_\mu^\text{A}$, we first observe that the obtained values lie visibly below recent estimates. This naturally results from an underestimation of the magnitude of the TFFs. A significant sensitivity to the AV meson mass is also observed. It is therefore clear that a precise description of the AV mesons, starting with their masses, is essential. From the perspective of the DS and BS equations, this is already a highly challenging task. Nonetheless, CI-based studies are valuable, as they provide a clear and transparent way to disentangle the role of each component, while maintaining the symmetry requirements and key non-perturbative features of the theory. With the TFFs rescaled according to the empirical estimate of the two-photon to $f_1$ decay, manifesting a conservative uncertainty, our final result aligns with recent estimates. This once again highlights that the main problem in the present analysis is the normalization of the produced TFFs. As we have observed, this could be achieved in a consistent scheme that simultaneously provides a good description of vector and axial-vector channels. This will be addressed in a subsequent work. Within this same framework, we will also study the process $\gamma^* \pi \to \gamma \gamma$, which plays a role in the data-driven determination of $a_\mu$.

\section*{Acknowledgements}
Work supported by: National Natural Science Foundation of China (Grant No. 12135007); Spanish Ministry of Science and Innovation (MICINN grant no.\ PID2022-140440NB-C22); and Junta de Andalucía (grant no.\ P18-FR-5057).

\appendix

\section{Alternate decomposition}
\label{ap:altbasis}
The amplitude decomposition for $\mathcal{A}_{\mu\nu\rho}$ suggested in Ref.\,\cite{Eichmann:2024glq} reads:
\begin{equation}
    \mathcal{A}_{\mu\nu\rho}(Q,Q')=m_{\text{A}}\sum_{i=1}^6\tau^i_{\mu\nu\rho}(Q,Q') F_i(Q^2,Q'^2)\,,
\end{equation}
where
\begin{align}
   m_{\text{A}}^3\, \tau_{\mu\nu\rho}^1&=\frac{1}{2}(\epsilon_{\mu\nu \alpha Q}\delta_{\alpha \nu}^{T}(Q')-\delta_{\mu\alpha}^{T}(Q)\epsilon_{\nu\rho \alpha Q'})\,,\nn
   m_{\text{A}}^5\, \tau_{\mu\nu\rho}^2&=\frac{\omega}{2}(\epsilon_{\mu\nu \alpha Q}\delta_{\alpha \nu}^{T}(Q')+\delta_{\mu\alpha}^{T}(Q)\epsilon_{\nu\rho \alpha Q'})\,,\nn
   m_{\text{A}}^5\, \tau_{\mu\nu\rho}^3&=2\omega \,\epsilon_{\mu\nu Q Q'}\Sigma_\rho\,,\nn
   m_{\text{A}}^3\, \tau_{\mu\nu\rho}^4&=\epsilon_{\mu\nu Q Q'} P_\rho\,,\nn
   m_{\text{A}}\, \tau_{\mu\nu\rho}^5&=\epsilon_{\mu\nu \rho \Sigma}\,,\nn
   m_{\text{A}}^3\, \tau_{\mu\nu\rho}^6&=\omega\,\epsilon_{\mu\nu \rho P}\,,
\end{align}
 with $\delta^{T}_{\mu\nu}(X)=X^2 \,\delta_{\mu\nu}-X_\mu X_\nu$ and $\Sigma=(Q+Q')/2$. The transversality of the AV meson and gauge invariance reduce the number of non-trivial scalar functions $F_i(Q^2,Q'^2)$. In particular, $F_{4,5,6}=0$ for all kinematics.

\section{Symmetry preserving regularization}
\label{ap:regu}

\subsection{Loop Integrals}
\label{ap:loopregu}
Due to the non-renormalizable nature of the CI, any loop integral must be regularized in a Poincare invariant manner. Herein we adopt the symmetry preserving regularization scheme described in Ref.~\cite{Xing:2022jtt}, which expresses any one loop integral with a class of the so-called irreducible loop integrals (ILIs) introduced in Ref.\,\cite{Wu:2002xa}:
\begin{align}
I_{-2\alpha}(\mathcal{M}^2)&=\int_{q}\frac{1}{(q^2+\mathcal{M}^2)^{\alpha+2}}\,,\nn
I^{\mu\nu}_{-2\alpha}(\mathcal{M}^2)&=\int_{q}\frac{q_{\mu}q_{\nu}}{(q^2+\mathcal{M}^2)^{\alpha+3}}\,,\nn
I^{\mu\nu\rho\sigma}_{-2\alpha}(\mathcal{M}^2)&=\int_{q}\frac{q_{\mu}q_{\nu}q_{\rho}q_{\sigma}}{(q^2+\mathcal{M}^2)^{\alpha+4}}\,.
\end{align}
These ILIs are regularized through the Schwinger's proper time method. For the scalar type ILIs
\begin{align}\label{eqn:reg}
I_{-2\alpha R}(\mathcal{M}^2) =\frac{1}{16\pi^2}\frac{\Gamma[\alpha,\tau_{uv}^2\mathcal{M}^2]-\Gamma[\alpha,\tau_{ir}^2\mathcal{M}^2]}{\mathcal{M}^{2\alpha}\Gamma(\alpha+2)}\,,
\end{align}
where $\tau_{uv,ir}$ are ultraviolet (UV) and infrared (IR) regulators, respectively, and $\Gamma(n,z)$ is the incomplete gamma function. 
The label $R$ stands for regularized integrals and will be suppressed for simplicity. For the tensor type ILIs, it is proved that they can be related with the scalar type ones through:
\begin{align}
I^{\mu\nu}_{-2\alpha R}(\mathcal{M}^2)&=\frac{\Gamma(\alpha+2)}{2\Gamma(\alpha+3)}\delta_{\mu\nu}I_{-2\alpha R}(\mathcal{M}^2)\,,\\
I^{\mu\nu\rho\sigma}_{-2\alpha R}(\mathcal{M}^2)&=\frac{\Gamma(\alpha+2)}{4\Gamma(\alpha+4)}S_{\mu\nu\rho\sigma}I_{-2\alpha R}(\mathcal{M}^2)\,.
\end{align}
These relations are precisely the so-called consistency conditions of gauge symmetry in Ref.~\cite{Wu:2002xa,Wu:2003dd}, which are independent of regularization and are necessary for preserving the gauge invariance of theories.


\subsection{Transition amplitude: Scalar functions}
\label{sec:TransAmp}
The structure of $\mathcal{A}_{\mu\nu\rho}$ demands the evaluation of a chiral trace, \emph{i.e.} that containing a $\gamma_5$ matrix. Regularizing such traces entails well-known subtleties\,\cite{Korner:1991sx,Jegerlehner:2000dz,Belusca-Maito:2023wah}. Herein we adopt the process described in Ref.\,\cite{Xing:2024bpj} to handle the chiral trace and express it in terms of scalar ILIs. For this purpose, the transition amplitude is first regularized employing dimensional regularization within the Breitenlohner-Maison/’t Hooft-Veltman scheme\,\cite{tHooft:1972tcz,Breitenlohner:1975hg}. Among others, this process avoids ambiguities from the Schouten Identity. The dimensionally regularized scalar ILIs are then mapped onto their counterparts in proper-time regularization. The resulting form factors can be expressed as follows:
\begin{align}
f_i=2N_{cf}E_{\text{A}}\sum_{X,Y}f_i^{XY}(Q^2,Q'^2)\hat{f}_X(Q^2)\hat{f}_Y(Q'^2)\,,
\end{align}
with $X\,,Y=\{L\,,T\,,A\}$ and $i=1,...,8$. The functions $\hat{f}_{X,Y}$ are those that appear in the QPV, Eq.\,\eqref{eq:QPVstru}. Meanwhile, after regularization, the functions $f_i^{XY}$ can be written as:
\begin{widetext}
\begin{subequations}
\begin{align}
    f_2^{TT'}&=\int_{\Omega}-4 ((3 u_2-1) I_0(\omega)+\mathcal{A}\, I_{-2}(\omega) )\,,\\
    f_2^{TA'}&=\int_{\Omega}8 (2 u_1 u_2 (m_\text{A}^2+Q^2+Q'^2)+Q^2 u_1 (2 u_1-3)+2 Q'^2 u_2^2-Q'^2 u_2)I_{-2}(\omega)\,,\\
    f_2^{AT'}&=\int_{\Omega}8 (u_2 (m_\text{A}^2+Q^2+Q'^2)-Q^2)I_{-2}(\omega)\,,\\
    f_2^{AA'}&=\int_{\Omega}4( Q^2 u_1 I_0(\omega)-\mathcal{B}\,I_{-2}(\omega) )/M^2\,,\\
    f_3^{LT'}&=\int_{\Omega}-4 ((3 u_2-1) I_0(\omega)+\mathcal{C}\, I_{-2}(\omega) )/Q^2\,,\\
    f_3^{LA'}&=\int_{\Omega}8 (Q^2 (1-2 u_1) u_1+Q'^2 u_2 (2 u_2-1))I_{-2}(\omega) /Q^2\,,\\
    f_3^{TT'}&=\int_{\Omega}4 ((3 u_2-1) I_0(\omega)+\mathcal{D}\,I_{-2}(\omega) )/Q^2\,,\\
    f_3^{TA'}&=\int_{\Omega}8 (Q^2 (3-2 u_1) u_1+Q'^2 (1-2 u_2) u_2)I_{-2}(\omega) /Q^2\,,\\
    f_3^{AT'}&=\int_{\Omega}8 I_{-2}(\omega)\,,\\
    f_3^{AA'}&=\int_{\Omega}-4( u_1 I_0(\omega)-\mathcal{E}\,I_{-2}(\omega))/M^2\,,\\
    f_4^{TT'}&=\int_{\Omega}-16 u_1 u_2 I_{-2}(\omega)\,,\\
    f_4^{TA'}&=\int_{\Omega}-32 u_1 u_2 I_{-2}(\omega)\,,\\
    f_4^{AT'}&=\int_{\Omega}-16 u_2 I_{-2}(\omega)\,,\\
    f_4^{AA'}&=\int_{\Omega}16 u_2 (-2 M^2+u_1 u_2 (m_\text{A}^2+Q'^2)+Q^2 u_1 (2 u_1+u_2-1)) I_{-2}(\omega)/M^2\,,\\
    f_7^{TT'}&=\int_{\Omega}-4 ( (3 u_1+3 u_2-2)I_0(\omega)+\mathcal{F}\,I_{-2}(\omega) )/m_\text{A}^2\,,\\
    f_7^{TA'}&=\int_{\Omega}8 (m_\text{A}^2 u_1 (2 u_2+1)+2 Q^2 u_1 (u_1+u_2-1)+Q'^2 (2 u_2+1) (u_1+u_2-1))I_{-2}(\omega)/m_\text{A}^2\,,\\
    f_7^{AT'}&=\int_{\Omega}8 (m_\text{A}^2 (2 u_1+1) u_2+Q^2 (2 u_1+1) (u_1+u_2-1)+2 Q'^2 u_2 (u_1+u_2-1))I_{-2}(\omega)/m_\text{A}^2\,,\\
    f_7^{AA'}&=\int_{\Omega}4 ( (Q^2 u_1+Q'^2 u_2)I_0(\omega)+\mathcal{G}\,I_{-2}(\omega))/(M^2 m_\text{A}^2)\,,
\end{align}
\end{subequations}

where $\int_\Omega=\int_0^1du_1\int_{0}^{1-u1}du_2=\int_0^1du_2\int_{0}^{1-u2}du_1$, $\omega=M^2-m_\text{A}^2 u_1 u_2-(u_1+u_2-1)(Q^2 u_1 +Q'^2 u_2)$ and 

\begin{subequations}
\begin{align}
    \mathcal{A}&=2 ((u_2-1) u_2 (2 m_\text{A}^2 u_1+Q'^2 (2 u_1+2 u_2-1))+Q^2 u_1 (2 u_1 (u_2-1)+u_2 (2 u_2-3)+2))\,,\\
    \mathcal{B}&=2(u_1 (Q^2 u_2 (2 m_\text{A}^2 (u_1+u_2-1)+Q'^2 (2 u_1-1))+u_2^2 (m_\text{A}^2+Q'^2)^2\nn&+Q^4 (2 u_1^2+u_1 (2 u_2-3)+(u_2-1)^2))-2 M^2 (u_2 (m_\text{A}^2+Q^2+Q'^2)-Q^2))\,,\\
    \mathcal{C}&=2 u_2  (m_\text{A}^2 u_1 (2 u_2-1)+2 Q^2 u_1 (u_1+u_2-1)+Q'^2 (2 u_2-1) (u_1+u_2-1))\,,\\
    \mathcal{D}&=2 (u_2 (2 u_2-1) (m_\text{A}^2 u_1+Q'^2 (u_1+u_2-1))+2 Q^2 u_1 (u_1 (u_2-1)+u_2 (u_2-1)+1))\,,\\
    \mathcal{E}&=2 (2 M^2+u_1 (-u_2 (m_\text{A}^2+2 Q'^2 u_2)+Q^2 u_1 (2 u_1-3)-Q^2 u_2+Q^2))\,,\\
    \mathcal{F}&=2 (2 m_\text{A}^2 u_1 u_2 (u_1+u_2-2)+Q^2 u_1 (u_1+u_2-1) (2 u_1+2 u_2-3)+Q'^2 u_2 (u_1+u_2-1) (2 u_1+2 u_2-3))\,,\\
    \mathcal{G}&=2  (2 M^2 (m_\text{A}^2 (u_1+u_2)+Q^2 (u_1+u_2-1)+Q'^2 (u_1+u_2-1))-2 Q^2 u_1 u_2 (m_\text{A}^2 (2 u_1+u_2-1)\nn&+Q'^2 (u_1+u_2-1))-u_2 (m_\text{A}^4 u_1 (u_1+u_2)+2 m_\text{A}^2 Q'^2 u_1 (u_1+2 u_2-1)\nn&+Q'^4 (u_1+u_2-1) (u_1+2 u_2-1))-Q^4 u_1 (u_1+u_2-1) (2 u_1+u_2-1))\,.
\end{align}
\end{subequations}
\end{widetext}
For the rest scalar functions, considering the symmetry of the Feynman parameters, \emph{i.e.} that the interchange $u_1\leftrightarrow u_2$ does not effect the Feynman parameter integration, one finds:
\begin{subequations}
\begin{align}
   f_1(Q^2,Q'^2)&=f_2(Q'^2,Q^2)\,,\\
    f_5(Q^2,Q'^2)&=f_4(Q'^2,Q^2)\,,\\
    f_6(Q^2,Q'^2)&=f_3(Q'^2,Q^2)\,,\\
    f_7(Q^2,Q'^2)&=f_7(Q'^2,Q^2)\,.
\end{align}
\end{subequations}
In addition, independent of the choice of $\xi$, one finds the relation
\begin{equation}
    f_8(Q^2,Q'^2)=-f_7(Q^2,Q'^2)\,,
\end{equation}
and if one turns off the AMM channel, \ie, $\xi=0$, one obtains
\begin{equation}
    f_4(Q^2,Q'^2)=f_5(Q^2,Q'^2)\,,
\end{equation}
which leads to $F_3(Q^2,Q'^2)=0$. That is, the form factor $F_3(Q^2,Q'^2)$ is generated by the AMM effect.

With the above expressions, one can demonstrate the gauge invariance of the amplitude $\mathcal{A}_{\mu\nu\rho}$ by showing that the scalar functions fulfill the relations in \Eqn{eq:wtis1}-\Eqn{eq:wtis2}. This verification is straightforward and the key steps given below.

\subsection{Transition amplitude: Gauge invariance}
\label{sec:GaugeInv}
To prove these relations, consider the following integral
\begin{equation}
    \mathcal{I}_1=\int_\Omega \partial_{u_2}(4 u_2 I_0(\omega))-\partial_{u_1}(4 u_1 I_0(\omega))
\end{equation}
treat the undifferentiated Feynman parameter as constant, one obtains
\begin{align}
    \mathcal{I}_1&=\int_0^1du_1(4u_2 I_0(\omega)|_{u_2=1-u_1}-4u_2 I_0(\omega)|_{u_2=0})\nn&-\int_0^1du_2(4u_1 I_0(\omega)|_{u_1=1-u_2}-4u_1 I_0(\omega)|_{u_1=0})\nn
    &=\int_0^1du_1(4(1-u_1)I_0(M^2-m_\text{A}^2u_1(1-u_1))\nn&-\int_0^1du_2(4(1-u_2)I_0(M^2-m_\text{A}^2u_2(1-u_2))
    \nn&=0
\end{align}
While directly computing the partial derivative:
\begin{align}
\mathcal{I}_1&=\int_\Omega4(I_0(\omega)+u_2\partial_{u_2}I_0(\omega)-I_0(\omega)-u_1\partial_{u_1}I_0(\omega))\nn&
    =\int_\Omega8(u_1 (1-2u_1)Q^2-u_2(1-2u_2)Q'^2 )I_{-2}(\omega)\,.
\end{align}
For the evaluation of the last line, we use the relation :
\begin{equation}
    I_{-2(\alpha+1)}(\mathcal{M}^2)=-\frac{1}{(\alpha+2)}\frac{d}{d\mathcal{M}^2}I_{-2\alpha}(\mathcal{M}^2)\,,
\end{equation}
which is valid under the proper time regularization \Eqn{eqn:reg}.
Therefore, one has:
\begin{equation}\label{eqn::iden1}
    0=\int_\Omega8(u_1 (1-2u_1)Q^2-u_2(1-2u_2)Q'^2 )I_{-2}(\omega)
\end{equation}

Similarly, by considering
\begin{equation}
    \mathcal{I}_2=\int_\Omega \partial_{u_2}(4(1-u_2) u_2 I_0(\omega))-\partial_{u_1}(4 u_1 u_2 I_0(\omega))\,,
\end{equation}
one arrives at another identity:
\begin{equation}\label{eqn::iden2}
    0=\mathcal{I}_2=\int_\Omega4 ((1-3 u_2) I_0(\omega)-\mathcal{C}\, I_{-2}(\omega))\,.
\end{equation}
Note that \Eqn{eqn::iden1} and \Eqn{eqn::iden2} leads to vannishing $f_3^{LT'}$ and $f_3^{LA'}$, which  indicates that the longitudinal part of the QPV does not contribute to the transition amplitude, as expected.
Pluggin in the regularized form factors $f_i$ into \Eqn{eq:wtis1}-\Eqn{eq:wtis2}, one finds that \Eqn{eq:wtis1} becomes:
\begin{equation}
    2N_{cf}E_{\text{A}}(  f_L(Q^2)f_T(Q'^2) \mathcal{I}_2 +  f_L(Q^2) f_A(Q'^2) \mathcal{I}_1)=0\,.
\end{equation}
On its part, \Eqn{eq:wtis1e} can be verified by just changing $Q^2$ and $Q'^2$ in \Eqn{eq:wtis1}. For the remaining piece, \Eqn{eq:wtis2}, we see that it trivially vanishes when inserting the form factors $f_i$ into it. This of course can be expected from the transverse structure of the AV meson BSA. At this stage, we have proved that the gauge invariance of the transition amplitude. It should be noted that, within our formalism and regularization, gauge invariance is preserved for arbitrary quark masses, AV meson masses, and the dressing functions of the QPV, even if these functions are not obtained from the corresponding equations.

\section{Parametrizations}
\label{ap:Params}

Let us consider the representation for the AV TFFs introduced in Ref.\,\cite{Eichmann:2024glq}:
\begin{equation}
    \label{eq:Param}
    F_i^\text{A}(x,w)=G_i^\text{A}(x)H_i^{\text{A}}(x,w)\,.
\end{equation}
Here $x=(Q^2+Q'^2)/(2m_\text{V})$ and $w=(Q^2-Q'^2)/(Q^2+Q'^2)$, with $|w|\leq1$. Cleary, $w=0$ corresponds to the symmetric limit, whereas $w=\pm 1$ stem from either $Q^2=0$ or $Q'^2=0$. The functions $G_i^{\text{A}}(x)$ and $H_i^\text{A}(x,w)$ are defined as follows:
\begin{eqnarray}
    G_i^\text{A}(x)&=&\frac{1}{(1+x)^{2\nu_i}}\left(a_i+b_i\,x + c_i x^{\nu_i}\frac{x+d_i}{x+e_i} \right)\,,\nonumber\\
    H_1^{A}(x,w)&=&1+\sum_{n=1}^{N}\frac{x^{\mu_1}}{x^{\mu_1}+\lambda_1n} \frac{w^{2n}}{1+\frac{2n}{3}}\,,\nonumber\\
    H_2^{A}(x,w)&=&1+\sum_{n=1}^{N}\frac{x^{\mu_2}}{x^{\mu_2}+\lambda_2n} \frac{(1+n) w^{2n}}{(1+\frac{2n}{3})(1+\frac{2n}{5})}\,,\nonumber\\
    H_3^\text{A}(x,w)&=&1\,.
\end{eqnarray}
The profile of $G_i^{\text{A}}(x)$ governs the power-law falloff, so we adopt $\nu_1=2$ and $\nu_{2,3}=3$. The functions $H_i^\text{A}(x,w)$ encode the asymptotic logarithms in the limit $N \to \infty$. For convenience, we set $N=100$. Recall, in our case, $m_{\rho,\phi}=0.879,\,1.058$ GeV. The rest of the parameters are collected in Table\,\ref{tab:Params1}.

\begin{table}[t]
\centering
\caption{Fit parameters for the $\{a_1,f_1,f_1'\}$ components of the AV TFFs, according to the representation in Eq.\,\eqref{eq:Param}.}
\label{tab:Params1}
\begin{tabular}{c||c|c|c||c|c|c}
 \multicolumn{7}{c}{\textbf{Case-I}}\\
\hline
 & $F_1^{a_1+f_1}$ & $F_2^{a_1+f_1}$ & $F_3^{a_1+f_1}$ & $F_1^{f_1'}$ & $F_2^{f_1'}$ & $F_3^{f_1'}$ \\
\hline
$a_i$ & 0.486 & 0.249 & -0.005 & 0.068 & 0.031 & -0.001 \\
$b_i$ & 0.955 & 0.944 & -0.016 & 0.124 & 0.127 & -0.003  \\
$c_i$ & 4.344 & 0.437 & -0.013 & 0.136 & 0.055 & -0.003  \\
$d_i$ & 17.800 & 2.363 & 0.797 & 11.317 & 3.082 & 1.414  \\
$e_i$ & 174.01 & 0.0 & 0.0 & 25.971 & 0.442 & 0.117  \\
$\mu_i$ & 1.434 & 1.552 & - & 1.396 & 1.464 & - \\
$\lambda_i$ & 4.235 & 5.813 & - & 4.553 & 6.331 & - \\
\hline
 \multicolumn{7}{c}{\textbf{Case-II}}\\
\hline
 & $F_1^{a_1+f_1}$ & $F_2^{a_1+f_1}$ & $F_3^{a_1+f_1}$ & $F_1^{f_1'}$ & $F_2^{f_1'}$ & $F_3^{f_1'}$ \\
\hline
$a_i$ & 0.465 & 0.231 & -0.003 & 0.067 & 0.032 & -0.001 \\
$b_i$ & 0.916 & 0.802 & -0.010 & 0.123 & 0.100 & -0.001  \\
$c_i$ & 8.733 & 0.443 & -0.098 & 0.136 & 0.057 & -0.002  \\
$d_i$ & 14.378 & 2.131 & 1.055 & 8.698 & 2.318 & 2.934  \\
$e_i$ & 296.089 & 0.0 & 0.188 & 19.650 & 0.0 & 0.0  \\
$\mu_i$ & 1.498 & 1.598 & - & 1.48 & 1.524 & - \\
$\lambda_i$ & 3.916 & 5.711 & - & 4.619 & 5.752 & - \\
\hline
\multicolumn{7}{c}{\textbf{Case-O}}\\
\hline
 & $F_1^{a_1+f_1}$ & $F_2^{a_1+f_1}$ & $F_3^{a_1+f_1}$ & $F_1^{f_1'}$ & $F_2^{f_1'}$ & $F_3^{f_1'}$ \\
\hline
$a_i$ & 0.448 & 0.196 & -0.002 & 0.070 & 0.032 & -0.001 \\
$b_i$ & 0.894 & 0.683 & -0.006 & 0.131 & 0.106 & -0.002  \\
$c_i$ & 2.886 & 0.425 & -0.173 & 0.141 & 0.070 & -0.002  \\
$d_i$ & 8.876 & 1.904 & 2.251 & 6.099 & 1.862 & 1.052  \\
$e_i$ & 59.960 & 0.0 & 57.629 & 12.685 & 0.0 & 0.301  \\
$\mu_i$ & 1.534 & 1.186 & - & 1.497 & 1.525 & - \\
$\lambda_i$ & 3.605 & 5.421 & - & 3.846 & 5.682 & - \\
\hline
\end{tabular}
\end{table}

\clearpage
\bibliographystyle{unsrt}
\bibliography{Evolving}

\end{document}